\documentclass[aip,jcp,reprint]{revtex4-1} 
\usepackage{graphicx}
\usepackage{enumitem}

\usepackage{amssymb}
\usepackage{mathtools}
\usepackage{graphicx}   
\usepackage{color}
\usepackage{ulem}
\usepackage{caption}
\usepackage{subcaption}
\usepackage{comment}

\usepackage{array}
\newcolumntype{C}[1]{>{\centering\arraybackslash}p{#1}} 
\usepackage{booktabs}
\usepackage{multirow}

\usepackage{booktabs}

\newcommand{\one}{\textcolor{black}}
\newcommand{\two}{\textcolor{black}}

\begin{document}

\title{Learning Molecular Dynamics with Simple Language Model built upon Long Short-Term Memory Neural Network}
\author{Sun-Ting Tsai}
\affiliation{Department of Physics and Institute for Physical Science and Technology, University of Maryland, College Park 20742, USA.}

\author{En-Jui Kuo}
\affiliation{Department of Physics and Joint Quantum Institute, University of Maryland, College Park 20742, USA.}

\author{Pratyush Tiwary*}
\email{ptiwary@umd.edu}
\affiliation{Department of Chemistry and Biochemistry and Institute for Physical Science and Technology,
 University of Maryland, College Park 20742, USA.}
\date{\today}

\begin{abstract}
Recurrent neural networks (RNNs) have led to breakthroughs in natural language processing and speech recognition, wherein hundreds of millions of people use such tools on a daily basis through smartphones, email servers and other avenues. In this work, we show such RNNs, specifically Long Short-Term Memory (LSTM) neural networks can also be applied to capturing the temporal evolution of typical trajectories arising in chemical and biological physics.  Specifically, we use a character-level language model based on LSTM. This learns a probabilistic model from 1-dimensional stochastic trajectories generated from molecular dynamics simulations of a higher dimensional system. We show that the model can not only capture the Boltzmann statistics of the system but it also reproduce kinetics at a large spectrum of timescales. We demonstrate how the embedding layer, introduced originally for representing the contextual meaning of words or characters, exhibits here a nontrivial connectivity between different metastable states in the underlying physical system. We demonstrate the reliability of our model and interpretations through different benchmark systems \one{and a single molecule force spectroscopy trajectory for multi-state riboswitch}. We anticipate that our work represents a stepping stone in the \one{understanding and }use of RNNs for modeling and predicting dynamics of complex stochastic molecular systems.

\end{abstract}

\maketitle

\section{Introduction}
Recurrent neural networks (RNN)\cite{jaeger2002tutorial} are a machine learning technique developed for modeling temporal sequences, with demonstrated successes including but not limited to modeling human languages.\cite{rico1992discrete,gicquel1998noninvertibility,graves2008novel,sundermeyer2012lstm,graves2013speech,sak2014long,cho2014learning,xingjian2015convolutional,chen2015lstm} A specific and extremely popular instance of RNNs are long short-term memory (LSTM)\cite{hochreiter1997long} neural networks, which possess more flexibility and can be used for challenging tasks such as language modeling, machine translation, and weather forecasting.\cite{sundermeyer2012lstm,luong2014addressing,xingjian2015convolutional} LSTMs were developed to alleviate the limitation of previously existing RNN architectures wherein they could not learn information originating from far past in time.  This is known as the vanishing gradient problem, a term that captures how the gradient or force experienced by the RNN parameters vanishes as a function of how long ago did the change happen in the underlying data.\cite{hochreiter2001gradient,agar2019revealing} LSTMs deal with this problem by controlling flows of gradients through a so-called gating mechanism where the gates can open or close determined by their values learned for each input. The gradients can now be preserved for longer sequences by deliberately gating out some of the effects. This way it has been shown that LSTMs can accumulate information for a long period of time by allowing the network to dynamically learn to forget aspects of information. Very recently LSTMs have also been shown to have the potential to mimic trajectories produced by experiments or simulations\cite{eslamibidgoli2019recurrent}, making accurate predictions about a short time into the future, given access to a large amount of data in the past. Similarly, another RNN variant named reservoir computing\cite{lukovsevivcius2009reservoir} has been recently applied to learn and predict chaotic systems.\cite{pathak2018model} Such a capability is already useful for instance in weather forecasting, where one needs extremely accurate predictions valid for a short period of time.

In this work, we consider an alternate and arguably novel use of RNNs, specifically LSTMs, in making predictions that in contrast to previous work \cite{pathak2018model,eslamibidgoli2019recurrent}, are valid for very long periods of time but only in a statistical sense. Unlike domains such as weather forecasting or speech recognition where LSTMs have allowed very accurate predictions albeit valid only for short duration of time, here we are interested in problems from chemical and biological physics, where the emphasis is more on making statistically valid predictions valid for extremely long duration of time. This is typified for example through the use of the ubiquitous notion of rate constant for activated barrier crossing, where short-time movements are typically treated as noise, and are not of interest for being captured through a dynamical model. Here we suggest an alternative way to use LSTM-based language model to learn a probabilistic model from the time sequence along some low-dimensional order parameters produced by computer simulations or experiments of a high-dimensional system. We also show by our computer simulations of different model systems that the language model can produce the correct Boltzmann statistics (as can other AI methods such as Ref. \cite{noe2019boltzmann}) but also the kinetics over a large spectrum of modes characterizing the dynamics in the underlying data. We highlight here a unique aspect of this calculation that the order parameter our framework needs could be arbitrarily close to or far from the true underlying slow mode, often called reaction coordinate. This in turn dictates how long of a memory kernel must be captured which is in general a very hard problem to solve.\cite{bussi2020using,wang2019past} Our framework is agnostic to proximity from the true reaction coordinate and reconstructs statistically accurate dynamics in a wide range of order parameters. Our work thus represents a new usage of a popular artificial intelligence (AI) framework to perform dynamical reconstruction in a domain of potentially high fundamental and practical relevance, including materials and drug design.

The manuscript is structured as follows: In Sec. \ref{sec:method} we explain the method and the neural network architecture we used in this work. In Sec. \ref{sec:pathentropy}, we show how the minimization of loss function leads to learning the path entropy of a physical system. In Sec. \ref{sec:embedding_kinetic}, we show the connection between the embedding layer and transition probability. Followed by this connection, we also show how we can define a transition probability through embedding vectors. Our computational results are then given in Sec. \ref{sec:computational_results}. The computational details including softwares we used are given in Sec. \ref{sec:testsystems}. \one{In Secs. \ref{sec:model_pots}, \ref{sec:aladi} and \ref{sec:sm_fret}, we shown our tests on Boltzmann statistics and kinetics for Langevin dynamics of model potentials, MD simulation of alanine dipeptide, and trajectory from single molecule force spectroscopy experiment on a multi-state riboswitch\cite{neupane2011single} respectively. In Sec. \ref{sec:embedding}, we compare numerically the transition probability introduced in Sec. \ref{sec:embedding_kinetic} with the actual counts in the trajectory.  In Sec. \ref{sec:Compare_MSM_HMM} as well as Supplementary Information (SI)  we compare our protocol with alternate approaches including Hidden Markov Models.}  Finally, a summary is given in Sec. \ref{se:conclusion}.

\section{Method}
\label{sec:method}

\begin{figure}[!h]
  \centering
  \includegraphics[width=8cm]{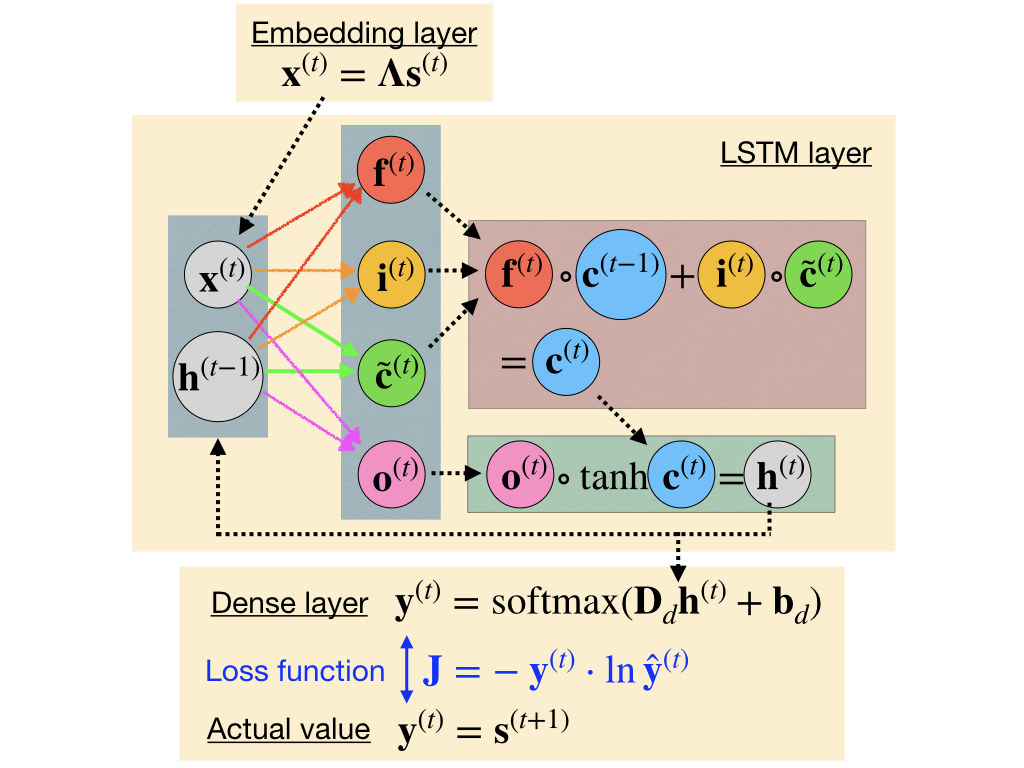}
  \caption
  {The schematic plot of the simple character-level language model used in this work. The model consists of three main parts: The embedding layer, the LSTM layer, and a dense output layer. The embedding layer is a linear layer which multiplies the one-hot input $\mathbf{s}^{(t)}$ by a matrix and produces an embedding vector $\mathbf{x}^{(t)}$. The $\mathbf{x}^{(t)}$ is then used as the input of LSTM network, in which the forget gate $\mathbf{f}^{(t)}$, the input gate $\mathbf{i}^{(t)}$, the output gate $\mathbf{o}^{(t)}$, and the candidate value $\tilde{\mathbf{c}}^{(t)}$ are all controlled by $(\mathbf{x}^{(t)}, \mathbf{h}^{(t-1)})$. The forget gate and input gate are then used to produce the update equation of cell state $\mathbf{c}^{t)}$. The output gate decides how much information propagates to the next time step. The output layer predicts the probabilities $\hat{\mathbf{y}}^{(t)}$ by parametrizing the transformation from $\mathbf{h}^{(t)}$ to $\hat{\mathbf{y}}$ with learned weights $\mathbf{D}_d$ and learned biases $\mathbf{b}_d$.  Finally, we can compute the cross entropy between the predicted probability distribution $\hat{\mathbf{y}}^{(t)}$ and the true probability distribution $\mathbf{y}^{(t)}=\mathbf{s}^{(t+1)}$. 
}\label{fig:model}
\end{figure}

Our central rationale in this work is that molecular dynamics (MD) trajectories, adequately discretized in space and time, can be mapped into a sequence of characters in some languages. By using a character-level language model that is effective in predicting future characters given the characters so far in a sequence, we can then learn the evolution of the MD trajectory that was mapped into the characters. The model we use is stochastic since it learns each character through the probability they appear in a corpus used for training. This language model consists of three sequential parts shown schematically in Fig. \ref{fig:model}. First, there is an embedding layer mapping one-hot vectors to dense vectors, followed by an LSTM layer which connects input states and hidden states at different time steps through a trainable recursive function, and finally a dense layer to transform the output of LSTM to the categorical probability vector.

Specifically, here we consider as input a one-dimensional time series produced by a physical system, for instance through Langevin dynamics being undergone by a complex molecular system. The time series consist of data points $\{\xi^{(t)}\}$, where $t$ labels the time step and $\xi\in\mathbb{R}$ is some one-dimensional collective variable or order parameter for the high-dimensional molecular system. In line with standard practice for probabilistic models, we convert the data points to one-hot encoded representations that implement spatial discretization. Thus each data point $\{\xi^{(t)}\}$ is represented by a $N$-dimensional binary vector $\mathbf{s}^{(t)}$, where $N$ is the number of discrete grid-points. An entry of one stands for the representative value and all the other entries are set to zeros. The representative values are in general finite if the order parameter is bounded, and are equally spaced in $\mathbb{R}$ with in total $N$ representative values. Note that the time series $\{\xi^{(t)}\}$ does not have to be one-dimensional. For a higher-dimensional series, we can always choose a set of representative values corresponding to locations in the higher-dimensional space visited trajectory. This would typically lead to a larger $N$ in the one-hot encoded representations, but the training set size itself will naturally stay the same. We find that the computational effort only depends on the size of training set and very weakly on $N$, and thus the time spent for learning a higher dimensional time series does not increase much relative to a one-dimensional series.

In the sense of modeling languages, the one-hot representation on its own cannot capture the relation between different characters. Take for instance that there is no word in the English language where the character $c$ is followed by $x$, unless of course one allows for the possibility of a space or some other letter in between. To deal with this, computational linguists make use of an embedding layer. The embedding layer works as a look-up table which converts each one-hot vector $\mathbf{s}^{(t)}$ to a dense vector $\mathbf{x}^{(t)}\in\mathbb{R}^M$ by the multiplication of a matrix $\mathbf{\Lambda}$ which is called the embedding matrix, where $M$ is called the embedding dimension
\begin{align}
    \mathbf{x}^{(t)}=\mathbf{\Lambda}\mathbf{s}^{(t)}
    \label{eq:embedding}
\end{align}

The sequence of dense representation $\mathbf{x}^{(t)}$ accounts for the relation between different characters as seen in the training time series. $\mathbf{x}^{(t)}$ is then used as the input of the LSTM layer. Each $\mathbf{x}^{(t)}$ generates an output $\mathbf{h}^{(t)}\in\mathbb{R}^L$ from LSTM layer, where $L$ is a tunable hyperparameter. Larger $L$ generally gives better learning capability but needs more computational resources. The LSTM itself consists of the following elements: the input gate $\mathbf{i}^{(t)}$, the forget gate $\mathbf{f}^{(t)}$, the output gate $\mathbf{o}^{(t)}$ the cell state $\mathbf{c}^{(t)}$, the candidate value $\tilde{\mathbf{c}}^{(t)}$, and $\mathbf{h}^{(t)}$ which is the hidden state vector and the final output from the LSTM. Each gate processes information in different aspects.\cite{hochreiter1997long} Briefly, the input gate decides which information to be written, the forget gate decides which information to be erased, and the output gate decides which information to be read from the cell state to the hidden state. The update equation of these elements can be written as follows:

\begin{align}
&\mathbf{f}^{(t)}=\sigma(\mathbf{W}_f\mathbf{x}^{(t)}+\mathbf{U}_f\mathbf{h}^{(t-1)}+\mathbf{b}_f) \label{eq:f_gate} \\
&\mathbf{i}^{(t)}=\sigma(\mathbf{W}_i\mathbf{x}^{(t)}+\mathbf{U}_i\mathbf{h}^{(t-1)}+\mathbf{b}_i) \label{eq:i_gate} \\
&\mathbf{o}^{(t)}=\sigma(\mathbf{W}_o\mathbf{x}^{(t)}+\mathbf{U}_o\mathbf{h}^{(t-1)}+\mathbf{b}_o) \label{eq:o_gate} \\
&\tilde{\mathbf{c}}^{(t)}=\tanh(\mathbf{W}_c\mathbf{x}^{(t)}+\mathbf{U}_c\mathbf{h}^{(t-1)}+\mathbf{b}_c) \label{eq:candidate} \\
&\mathbf{c}^{(t)}=\mathbf{f}^{(t)}\circ \mathbf{c}^{(t-1)}+\mathbf{i}^{(t)}\circ\tilde{\mathbf{c}}^{(t)} \label{eq:cell} \\
&\mathbf{h}^{(t)}=\mathbf{o}^{(t)}\circ\tanh(\mathbf{c}^{(t)}) \label{eq:hidden}
\end{align}
where $\mathbf{W}$ and $\mathbf{b}$ are the corresponding weight matrices and bias vectors. The $\tanh(\mathbf{v})$ operates piecewise on each element of the vector $\mathbf{v}$. The operation $\circ$ is the Hadamard product.\cite{goodfellow2016deep}

The final layer in Fig. \ref{fig:model} is a simple dense layer with fully connected neurons which converts the output $\mathbf{h}^{(t)}$ of the LSTM to a vector $\mathbf{y}^{(t)}$ in which each entry denotes the categorical probability of the representative value for the next time step $t+1$. The loss function $J$ for minimization during training at every timestep $t$ is then defined as the cross entropy between the output of the model $\hat{\mathbf{y}}^{(t)}$ and the actual probability for the next timestep $\hat{\mathbf{y}}^{(t)}$ which is just the one-hot vector $\mathbf{s}^{t+1}$
\begin{align}
    &\hat{\mathbf{y}}^{(t)}=\mathrm{softmax}(\mathbf{D}_{d}\mathbf{h}^{(t)}+\mathbf{b}_{d}) \label{eq:output} \\
    &J=-\sum^{T-1}_{t=0}\mathbf{y}^{(t)}\cdot\ln\hat{\mathbf{y}}^{(t)}=-\sum^{T-1}_{t=0}\mathbf{s}^{(t+1)}\cdot\ln\hat{\mathbf{y}}^{(t)} \label{eq:loss}
\end{align}
where $T$ is the total length of trajectory, and the final loss function is the sum over the whole time series. The $\mathrm{softmax(\mathbf{x})}_i=\exp(\mathbf{x}_i)/\sum_j \exp(\mathbf{x}_j)$ is a softmax function mapping $\mathbf{x}$ to a probability vector $\hat{\mathbf{y}}$.

\section{Theory}
\label{sec:theory}
\subsection{Training LSTM is equivalent to learning path entropy}
\label{sec:pathentropy}
The central finding of this work, which we demonstrate through numerical results for different systems in Sec. \ref{sec:computational_results}, is that a LSTM framework used to model languages can also be used to capture kinetic and thermodynamic aspects of dynamical trajectories prevalent in chemical and biological physics. In this section we demonstrate theoretically as to why LSTMs possess such a capability. Before we get into the mathematical reasoning detailed here as well as in SI, we first state our key conceptual idea. Minimizing the loss function $J$ in LSTM (Eq. \ref{eq:loss}), which trains the model at time $t$ to generate output $\hat{\mathbf{y}}^{(t)}$ resembling the target output $\mathbf{s}^{t+1}$, is equivalent to minimizing the difference between the actual and LSTM-learned path probabilities. This difference between path probabilities can be calculated as a cross-entropy $J'$ defined as: 
\begin{align}
    J'=-\sum_{\mathbf{x}^{(T)}... \mathbf{x}^{(0)}}P(\mathbf{x}^{(T)} ... \mathbf{x}^{(0)})\ln Q(\mathbf{x}^{(T)} ... \mathbf{x}^{(0)})
    \label{eq:cross_path_entropy}
\end{align}
where $P(\mathbf{x}^{(t+1)}, ..., \mathbf{x}^{(0)})$ and $Q(\mathbf{x}^{(t+1)}, ..., \mathbf{x}^{(0)})$ are the corresponding true and neural network learned path probabilities of the system. Eq. \ref{eq:cross_path_entropy} can be rewritten\cite{cover2012elements} as the sum of path entropy $H(P)$ for the true distribution $P$ and Kullback-Liebler distance $D_{KL}$ between $P$ and $Q$: $J' = H(P) + D_{KL}(P||Q)$. Since $D_{KL}$ is strictly non-negative\cite{cover2012elements} attaining the value of $0$ iff $Q=P$, the global minimum of $J'$ happens when $Q=P$ and $J'$ equals the path entropy $H(P)$ of the system.\cite{presse2013principles} Thus we claim that minimizing the loss function in LSTM is equivalent to learning the path entropy of the underlying physical model, which is what makes it capable of capturing kinetic information of the dynamical trajectory.

To prove this claim we start with rewriting $J$ in Eq. \ref{eq:loss}. For a long enough observation period $T$ or for a very large number of trajectories, $J$ can be expressed as the cross entropy between conditional probabilities:
\begin{align}
    J=-\sum^{T-1}_{t=0}\sum_{\mathbf{x}^{(t+1)}} &P(\mathbf{x}^{(t+1)}|\mathbf{x}^{(t)} ... \mathbf{x}^{(0)}) \nonumber \\
    &\times\ln Q(\mathbf{x}^{(t+1)}|\mathbf{x}^{(t)} ... \mathbf{x}^{(0)})
    \label{eq:J_equiv}
\end{align}
where $P(\mathbf{x}^{(t+1)}|\mathbf{x}^{(t)} ... \mathbf{x}^{(0)})$ is the true conditional probability for the physical system, and $Q(\mathbf{x}^{(t+1)}|\mathbf{x}^{(t)} ... \mathbf{x}^{(0)})$ is the conditional probability learned by the neural network. The minimization of Eq. 11 leads to minimization of the cross entropy $J'$ as shown in the SI. Here we conversely show how Eq. \ref{eq:cross_path_entropy} reduces to Eq. \ref{eq:loss} by assuming a stationary first-order Markov process as in Ref. \onlinecite{presse2013principles}:
\begin{align}
    P(\mathbf{x}^{(T)} ... \mathbf{x}^{(0)})&=P(\mathbf{x}^{(T)}|\mathbf{x}^{(T-1)})...P(\mathbf{x}^{(1)}|\mathbf{x}^{(0)})P(\mathbf{x}^{(0)})  \nonumber \\
    Q(\mathbf{x}^{(T)} ... \mathbf{x}^{(0)})&=Q(\mathbf{x}^{(T)}|\mathbf{x}^{(T-1)}) ... Q(\mathbf{x}^{(1)}|\mathbf{x}^{(0)})Q(\mathbf{x}^{(0)})
    \label{eq:markovprob}
\end{align}
where $P(\mathbf{x}^{(t+1)}_j|\mathbf{x}^{(t)}_i) \equiv P_{ij}$ is the transition probability from state $\mathbf{x}_i$ to state $\mathbf{x}_j$ and $P(\mathbf{x}^{(0)}_k)\equiv P_k$ is the occupation probability for the single state $\mathbf{x}_k$. Plugging Eq. \ref{eq:markovprob} into Eq. \ref{eq:cross_path_entropy}, and following the derivation in Ref. \onlinecite{presse2013principles} with the constraints
\begin{align}
    \sum_j P_{ij}=1 \quad \sum_i P_i P_{ij}=P_j
\label{eq:constraints}    
\end{align}
we arrive at an expression for the cross-entropy $J$, which is very similar to the path entropy type expressions derived for instance in the framework of Maximum Caliber\cite{presse2013principles}:
\begin{align}
    J'&=-\sum_i P_i\ln Q_i-T\sum_{lm}P_l P_{lm}\ln (Q_{lm}) \label{eq:J_after_constraints} \\
    &\to -T\sum_{lm}P(\mathbf{x}_l) P(\mathbf{x}_m|\mathbf{x}_l)\ln Q(\mathbf{x}_m|\mathbf{x}_l)
    \label{eq:J_after_constraints_limit}
\end{align}
In Eq. \ref{eq:J_after_constraints} as the trajectory length $T$ increases, the second term dominates in the estimate of $J$ leading to Eq. \ref{eq:J_after_constraints_limit}. This second term is the ensemble average of a time-dependent quantity $\tilde{J}(\mathbf{x}^{(t)}_l)\equiv-\sum_m P(\mathbf{x}^{(t+1)}_m|\mathbf{x}^{(t)}_l)\ln Q(\mathbf{x}^{(t+1)}_m|\mathbf{x}^{(t)}_l)$. For a large enough $T$, the ensemble average can be replaced by the time average. By assuming ergodicity\cite{moore2015ergodic}:
\begin{align}
    J'=-\sum^{T}_{t=1}\sum_m P(\mathbf{x}^{(t+1)}_m|\mathbf{x}^{(t)}_l)\ln Q(\mathbf{x}^{(t+1)}_m|\mathbf{x}^{(t)}_l)
    \label{eq:avg_cross_entropy}
\end{align}
from which we directly obtain Eq. \ref{eq:loss}. Therefore, under first-order Markovianity and ergodicity, minimizing the loss function $J$ of Eq. \ref{eq:loss} is equivalent to minimizing $J'$ and thereby learning the path entropy. In the SI we provide a proof for this statement that lifts the Markovianity assumption as well - the central idea there is similar to what we showed here.

\subsection{Embedding layer in LSTM captures kinetic distances}
\label{sec:embedding_kinetic}

In word embedding theory, the embedding layer provides a measure of similarity between words. However, from the path probability representation, it is unclear how the embedding layer works since the derivation can be done without embedding vectors $\mathbf{x}$. To have an understanding to $Q_{lm}$ in the first-order Markov process, 
we first write the conditional probability $Q_{lm}=Q(\mathbf{x}^{(t+1)}_m|\mathbf{x}^{(t)}_l)$ explicitly with softmax defined in Eq. \ref{eq:output} and embedding vectors $\mathbf{x}$ defined in Eq. \ref{eq:embedding}:
\begin{align}
    Q_{lm}&=\frac{\exp(\mathbf{s}^{(t+1)}_m\cdot(\mathbf{D}_d\mathbf{h}^{(t)}+\mathbf{b}_d))}{\sum_k\exp(\mathbf{s}_k\cdot(\mathbf{D}_d\mathbf{h}^{(t)}+\mathbf{b}_d))} \nonumber \\
    &=\frac{\exp(\mathbf{s}^{(t+1)}_m\cdot(\mathbf{D}_d f_{\boldsymbol{\theta}}(\mathbf{x}^{(t)})+\mathbf{b}_d))}{\sum_k\exp(\mathbf{s}_k\cdot (\mathbf{D}_d f_{\boldsymbol{\theta}}(\mathbf{x}^{(t)})+\mathbf{b}_d))}
    \label{eq:conditional_p}
\end{align}
where $f$ is the recursive function $\mathbf{h}^{(t)}=f_{\boldsymbol{\theta}}(\mathbf{x}^{(t)}, \mathbf{h}^{(t-1)})\approx f_{\boldsymbol{\theta}}(\mathbf{x}^{(t)})$ which is defined with the update equation in Eq. \ref{eq:f_gate}-\ref{eq:hidden}. In Eq. \ref{eq:conditional_p}, $\boldsymbol{\theta}$ denotes various parameters including all weight matrices and biases, and the summation index $k$ runs over all possible states. Now we can use multivariable Taylor's theorem to approximate $f_{\boldsymbol{\theta}}$ as the linear term around a point $\mathbf{a}$ as long as $\mathbf{a}$ is not at any local minimum of $f_{\boldsymbol{\theta}}$:
\begin{align}
    f_{\boldsymbol{\theta}}(\mathbf{x}^{(t)})\approx f_{\boldsymbol{\theta}}(\mathbf{a})+\mathbf{A}_{\boldsymbol{\theta}}(\mathbf{x}^{(t)}-\mathbf{a})
\end{align}
where $\mathbf{A}_{\boldsymbol{\theta}}$ is the $L$ by $M$ matrix defined to be $(\mathbf{A}_{\boldsymbol{\theta}})_{ij}=\frac{\partial (f_{\boldsymbol{\theta}})_i}{\partial x_j}|_{\mathbf{x}=\mathbf{a}}$. Then Eq. \ref{eq:conditional_p} becomes
\begin{align}
    Q_{lm}=\frac{\exp(C^{(t+1)}_m)\exp(\mathbf{s}^{(t+1)}_m\cdot\mathbf{D}_d\mathbf{A}_{\boldsymbol{\theta}}\mathbf{x}^{(t)}_l)}{ \sum_k\exp(C_k)\exp(\mathbf{s}_k\cdot\mathbf{D}_d\mathbf{A}_{\boldsymbol{\theta}}\mathbf{x}^{(t)}_l)}
    \label{eq:skipgramlikeQ}
\end{align}
where $C^{(t+1)}_i=\mathbf{s}^{(t+1)}_i\cdot[\mathbf{D}_d (f_{\boldsymbol{\theta}}(\mathbf{a}_l)+\mathbf{A}_{\boldsymbol{\theta}}\mathbf{a}_l)+\mathbf{b}_d]$. We can see in Eq. \ref{eq:skipgramlikeQ}  how the embedding vectors come into the transition probability. Specifically, there is a symmetric form between output one-hot vectors $\mathbf{s}^{(t+1)}_m$ and the input one-hot vectors $\mathbf{s}^{(t)}$, in which $\mathbf{x}^{(t)}=\boldsymbol{\Lambda}\mathbf{s}^{(t)}$ and $\boldsymbol{\Lambda}$ is the input embedding matrix, $\mathbf{D}_d\mathbf{A}_{\boldsymbol{\theta}}$ can be seen as the output embedding matrix, and $C^{(t+1)}_i$ is the correction of time lag effect.  While we don't have an explicit way to calculate the output embedding matrix so defined, Eq. \ref{eq:skipgramlikeQ} motivates us to define the following \textit{ansatz} for the transition probability:
\begin{align}
    Q_{lm}=Q(\mathbf{x}_m|\mathbf{x}_l)=\frac{\exp(\mathbf{x}_m\cdot\mathbf{x}_l)}{\sum_k\exp(\mathbf{x}_k\cdot\mathbf{x}_l)} \label{eq:conditional_p_fin}
\end{align}
where $\mathbf{x}_m$ and $\mathbf{x}_l$ are both calculated by the input embedding matrix $\boldsymbol{\Lambda}$. The expression in  Eq. \ref{eq:conditional_p_fin} is thus a tractable approximation to the more exact transition probability in Eq. \ref{eq:skipgramlikeQ}. Furthermore, we will show in Sec. \ref{sec:embedding} through numerical examples of test systems that our \textit{ansatz} for $Q_{lm}$ does correspond to the kinetic connectivity between states. That is, the LSTM embedding layer with the transition probability through Eq. \ref{eq:conditional_p_fin} can capture the average commute time between two states in the original physical system, irrespective of the quality of low-dimensional projection fed to the LSTM. \cite{noe2016commute,noe2015kinetic,tsai2020distance}

\section{Computational results}
\label{sec:computational_results}

\subsection{Test systems}
\label{sec:testsystems}

To demonstrate our ideas, here we consider \one{a range of different dynamical trajectories. These include three model potentials, the popular model molecule alanine dipeptide, and trajectory from single molecule force spectroscopy experiments on a multi-state riboswitch.\cite{neupane2011single}}  When applying our neural network to the model systems, the embedding dimension $M$ is set to 8 and LSTM unit $L$ set to 64. \one{When learning trajectories for alanine dipeptide and riboswitch}, we took $M=128$ and $L=1024$. All time series were batched into sequences with a sequence length of 100 and the batch size of 64. For each model potential, the neural network was trained using the method of stochastic gradient descent for 20 epochs until the training loss becomes smaller than the validation loss, which means an appropriate training has been reached. For alanine dipeptide, 40 training epochs  were used. Our neural network was built using TensorFlow version 1.10.

All model potentials have two degrees of freedom $x$ and $y$. Our first two models (shown in Fig. \ref{fig:Boltzmann_3s_4s}(a) and Fig. \ref{fig:Boltzmann_3s_4s}(b)) have three metastable states with governing potential $U(x,y)$ given by 
\begin{align}
    U(x, y)&=W(x^6+y^6)-G(x, x_1)G(y, y_1) \nonumber \\
    &-G(x, x_2)G(y, y_2)-G(x, x_3)G(y, y_3)
    \label{eq:firsttwo_pot}
\end{align}
where $W=0.0001$ and $G(x, x_0)=e^{-\frac{(x-x0)^2}{2\sigma^2}}$ denotes a Gaussian function centered at $x_0$ with width $\sigma=0.8$. We also build a 4-state model system with governing interaction potential:
\begin{align}
    U(x, y)&=W(x^4+y^4)+G(x, 0.0)G(y, 0.0) \nonumber \\
    &-G(x, 2.0)G(y, -1.0)-G(x, 0.5)G(y, 2.0) \nonumber \\
    &-G(x, -0.5)G(y, -2.0)-G(x, -2.0)G(y, 1.0)
    \label{eq:fourstate_pot}
\end{align}
The different local minima corresponding to the model potentials in Eq. \ref{eq:firsttwo_pot} and Eq. \ref{eq:fourstate_pot} are illustrated in Fig. \ref{fig:Boltzmann_3s_4s}. We call these as linear 3-state, triangular 3-state, and  4-state models respectively. The free energy surfaces generated from the simulation of Langevin dynamics\cite{bussi2007accurate} with these model potentials are shown in Figs. \ref{fig:Boltzmann_3s_4s}(a)-(c). The integration timestep for the Langevin dynamics simulation was 0.01 units, and the simulation was performed at $\beta=9.5$ for linear 3-state and 4-state potentials and  $\beta=9.0$ for triangular 3-state potential, where $\beta=1/k_BT$. The MD trajectory for alanine dipeptide was obtained using the software GROMACS 5.0.4\cite{berendsen1995gromacs,abraham2015gromacs}, patched with PLUMED 2.4\cite{plumed2019nature}. The temperature was kept constant at 450K using the velocity rescaling thermostat\cite{bussi2007canonical}.

\subsection{Boltzmann statistics and kinetics for model potentials}
\label{sec:model_pots}
\begin{figure}[!h]
  \centering
  \includegraphics[width=8cm]{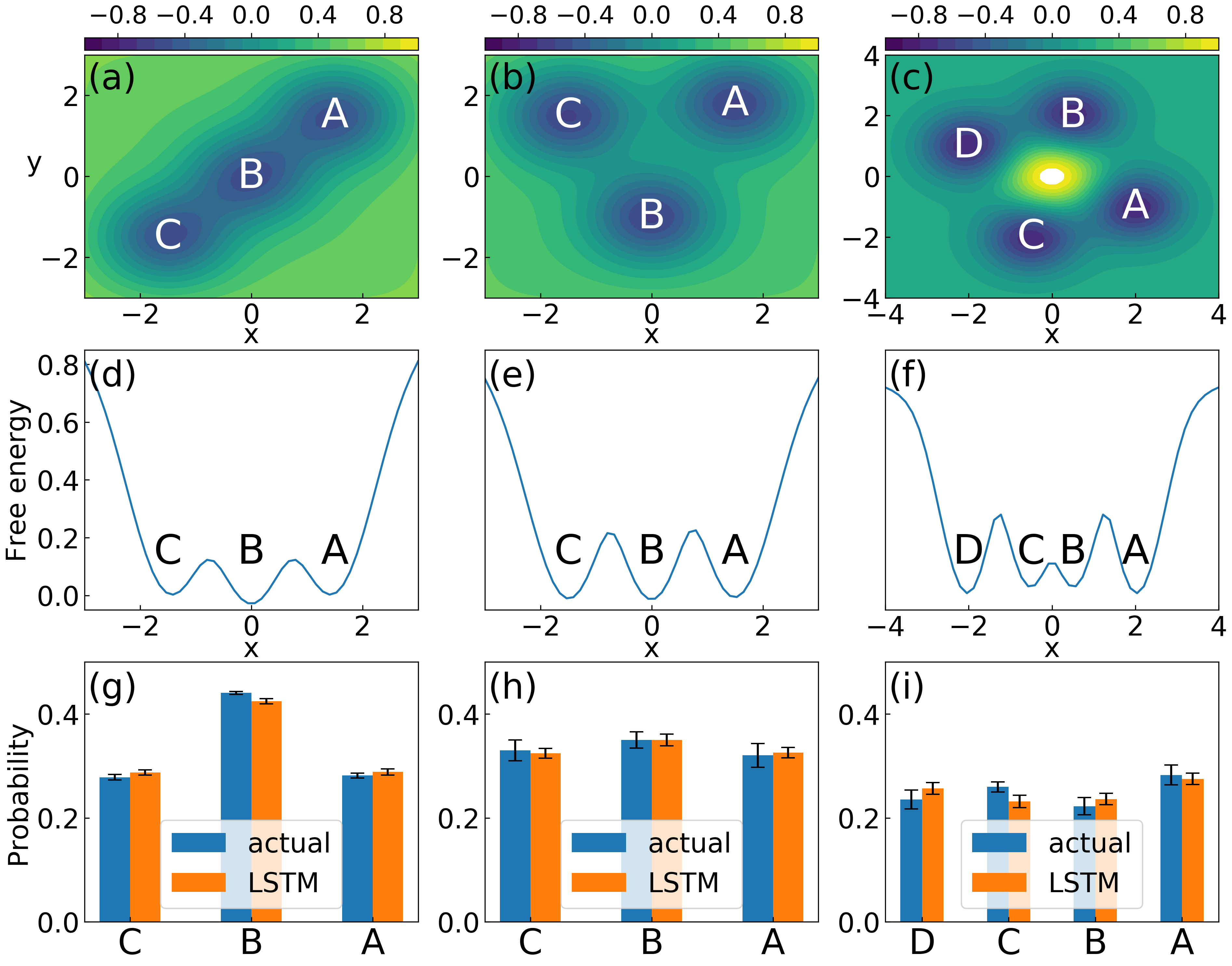}
  \caption
  {The analytical free energy generated from (a) linear 3-state, (b) triangular 3-state, (c) symmetric 4-state model potentials and (d), (e), (f) are the corresponding 1-dimensional projections along x-direction. In the bottom, we compare the Boltzmann probabilities of (g) linear 3-state, (h) triangular 3-state, and (i) symmetric 4-state models at each labeled states generated from actual MD simulation and from our neural network model.
}\label{fig:Boltzmann_3s_4s}
\end{figure}

The first test we perform for our model is its ability to capture the Boltzmann weighted statistics for the different states in each model potential. This is the probability distribution $P$ or equivalently the related free energy $F=-{1 \over \beta} \text{log} P$, and can be calculated by direct counting from the trajectory. As can be seen in Fig. \ref{fig:Boltzmann_3s_4s}, the LSTM does an excellent job of recovering the Boltzmann probability within error bars.

Next we describe our LSTM deals with a well-known problem in analyzing high-dimensional data sets through low-dimensional projections. One can project the high-dimensional data along many different possible low-dimensional order parameters, for instance $x$, $y$ or a combination thereof in Fig. \ref{fig:Boltzmann_3s_4s}. However most such projections will end up not being kinetically truthful and give a wrong impression of how distant the metastable states actually are from each other in the underlying high-dimensional space. It is in general hard to come up with a projection that preserves the kinetic properties of the high-dimensional space. Consequently, it is hard to design analysis or sampling methods that even when giving a time-series along a sub-optimal projection, still capture the true kinetic distance in the underlying high-dimensional space.

Here we show how our LSTM model is agnostic to the quality of the low-dimensional projection in capturing accurate kinetics. Given that for each of the 3 potentials the LSTM was provided only the $x-$trajectory, we can expect that the chosen model potentials constitute different levels of difficulties in generating correct kinetics. Specifically, a one-dimensional projection along $x$ is kinetically truthful for the linear 3-state potential in Fig. \ref{fig:Boltzmann_3s_4s}(a) but not for the triangular 3-state and the 4-state potentials in Figs. \ref{fig:Boltzmann_3s_4s}(b) and (c) respectively. For instance, Fig. \ref{fig:Boltzmann_3s_4s}(e) gives the impression that state C is kinetically very distant from state A, while in reality for this potential all 3 pairs of states are equally close to each other. Similar concerns apply to the 4-state potential.

\begin{figure}[!h]
  \centering
  \includegraphics[width=8cm]{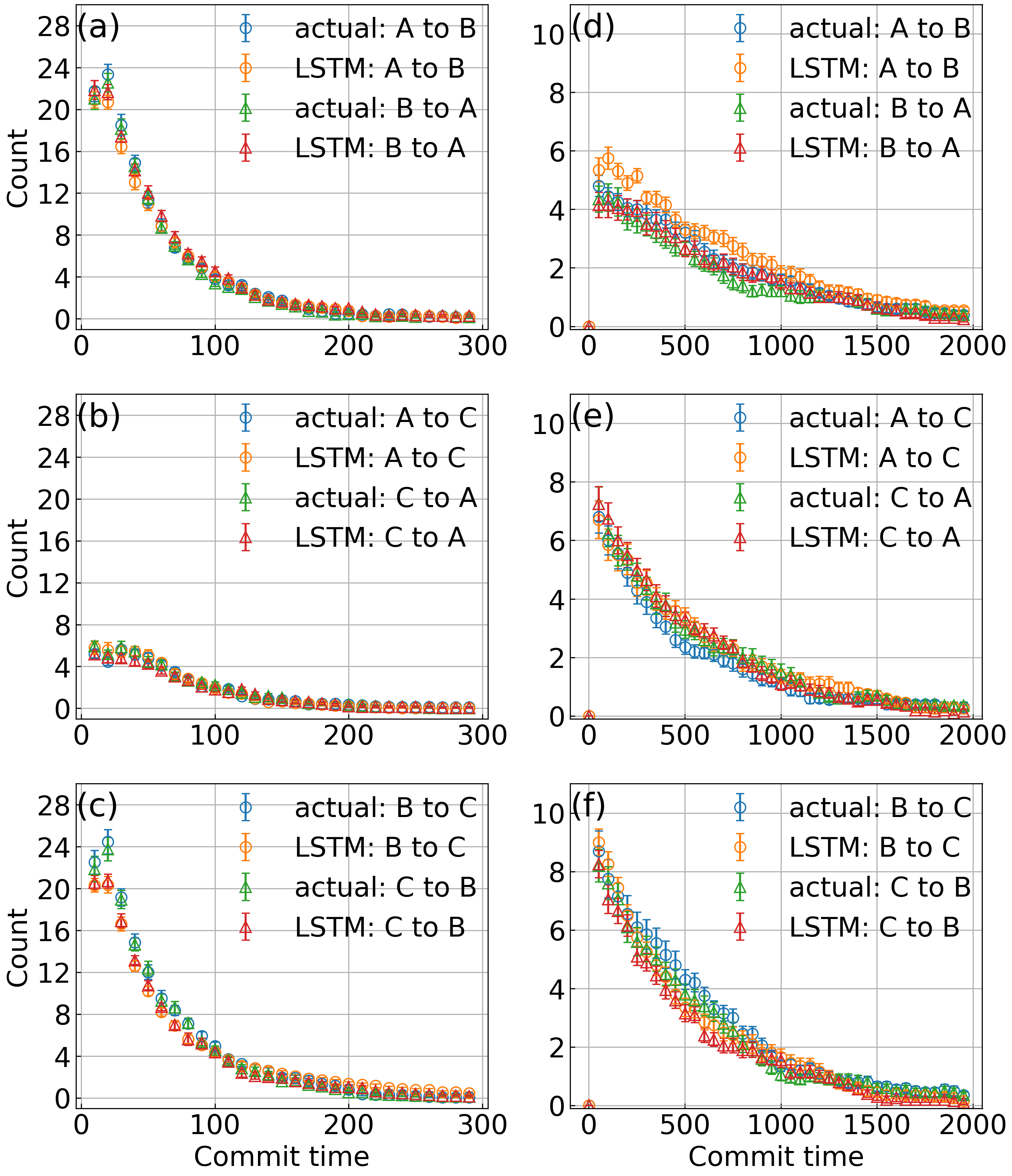}
  \caption
  {Number of transitions between different pairs of metastable states as a function of commitment time defined in Sec. \ref{sec:model_pots}. The calculations for linear and triangular configurations are shown in (a)-(c) and (d)-(f) respectively.
}\label{fig:Commit_t_3sTranspose}
\end{figure}

\begin{figure}[!h]
  \centering
  \includegraphics[width=8cm]{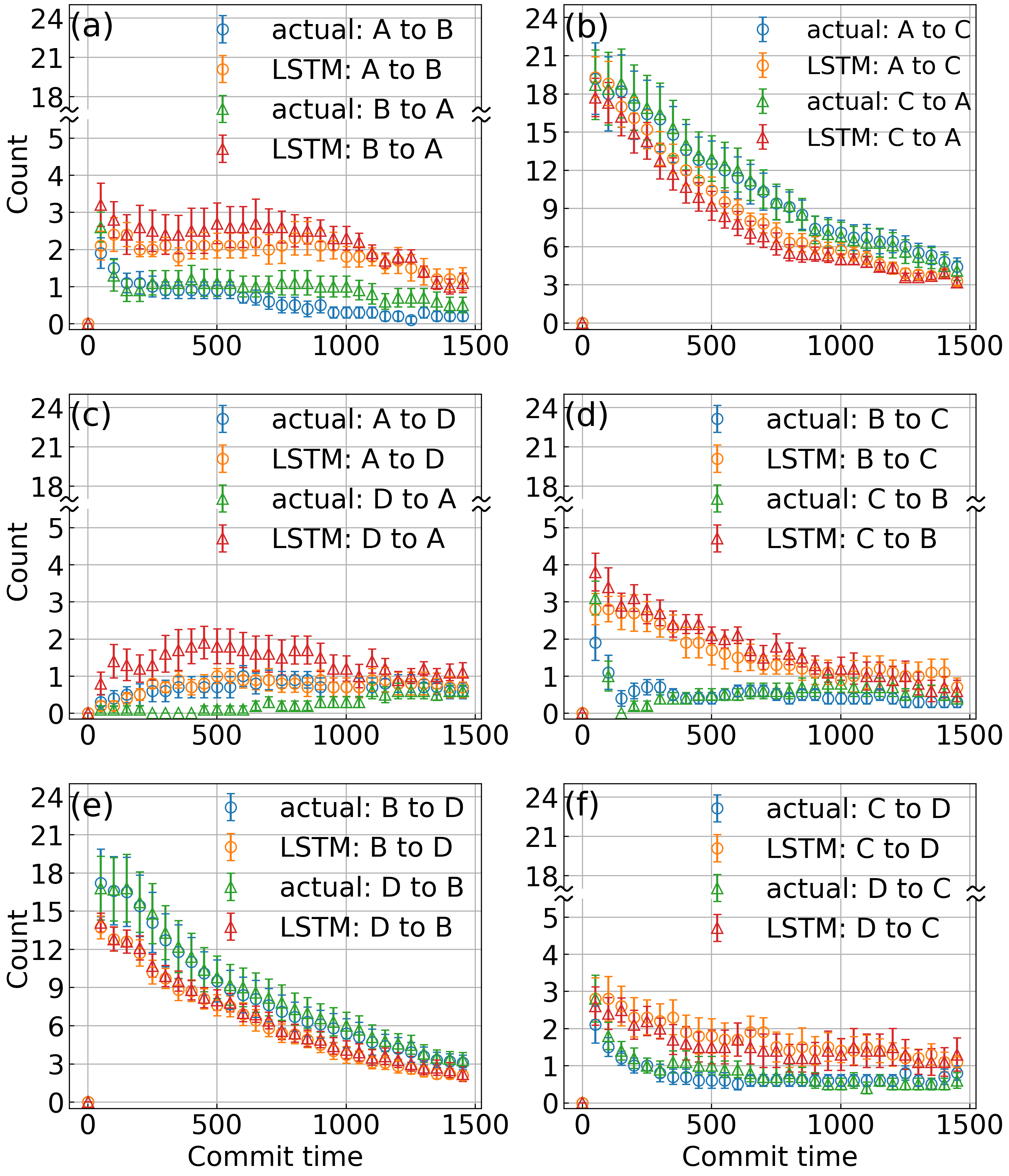}
  \caption
  {Number of transitions between different pairs of metastable states as a function of commitment time defined in Sec. \ref{sec:model_pots} for 4-state model system.
}\label{fig:Commit_t_4sTranspose}
\end{figure}

In Figs. \ref{fig:Commit_t_3sTranspose} and \ref{fig:Commit_t_4sTranspose} (a)-(c) and (d)-(f) we compare the actual versus LSTM-predicted kinetics for moving between different metastable states for different model potentials, for all pairs of transitions in both directions (i.e. for instance A to B and B to A). Specifically, Fig. \ref{fig:Commit_t_3sTranspose} (a)-(c) and Fig. \ref{fig:Commit_t_3sTranspose} (d)-(f) shows results for moving between the 3 pairs of states in the linear and triangular 3-state potentials respectively. Fig. \ref{fig:Commit_t_4sTranspose} shows results for the 6 pairs of states in the 4-state potential. Furthermore, for every pair of state, we analyze the transition time between those states as a function of different minimum commitment or commit time, i.e. the minimum time that must be spent by the trajectory in a given state to be classified as having committed to it. A limiting value, and more specifically the rate at which the population decays to attain to such a limiting value, corresponds to the inverse of the rate constant for moving between those states. \cite{hanggi1990reaction,berne1988classical}  Thus here we show how our LSTM captures not just the rate constant, but time-dependent fluctuations in the population in a given metastable state as equilibrium is attained. The results are averaged over 20 independent segments taken from the trajectories of different trials of training for the 3-state potentials and 10 independent segments for the 4-state potential.

As can be seen in Figs. \ref{fig:Commit_t_3sTranspose} and \ref{fig:Commit_t_4sTranspose}, the LSTM model does an excellent job of reproducing well within errorbars the transition times between different metastable states for different model potentials irrespective of the quality of the low-dimensional projection.
Firstly, our model does tell the differences between linear and triangular 3-state models (Fig. \ref{fig:Commit_t_3sTranspose}) even though the projected free energies along the $x$ variable input into LSTM are same (Fig. \ref{fig:Boltzmann_3s_4s}). The number of transitions between states A and C is less than the others; while for triangular configuration, the numbers of transitions between all pairs of states are similar. The rates at which the transition count decays as a function of commitment time is also preserved between the input data and the LSTM prediction.

The next part of our second test is the 4-state model potential. In Fig. \ref{fig:Commit_t_4sTranspose} we show comparisons for all 6 pairs of transitions in both forward and reverse directions. A few features are immediately striking here. Firstly, even though states B and C are perceived to be kinetically proximal from the free energy (Fig. \ref{fig:Boltzmann_3s_4s}), the LSTM captures that they are distal from each other and correctly assigns similar kinetic distance to the pairs B,C as it does to A,D. Secondly, there is asymmetry between the forward and backward directions (for e.g. A to D and D to A, indicating that the input trajectory itself has not yet sufficiently sampled the slow transitions in this potential. As can be seen from Fig. \ref{fig:Boltzmann_3s_4s} (c) the input trajectory has barely 1 or 2 direct transitions for the very high barrier A to D or B to C. This is a likely explanation for why our LSTM model does a bit worse than in the other two model potentials in capturing the slowest transition rates, as well as the higher error bars we see here. In other words, so far we can conclude that while our LSTM model can capture equilibrium probabilities and transition rates for different model potentials irrespective of the input projection direction or order parameter, it is still not a panacea for insufficient sampling itself, as one would expect.

\subsection{Boltzmann statistics and kinetics for alanine dipeptide}
\label{sec:aladi}
\begin{figure}[!h]
  \centering
  \includegraphics[width=8cm]{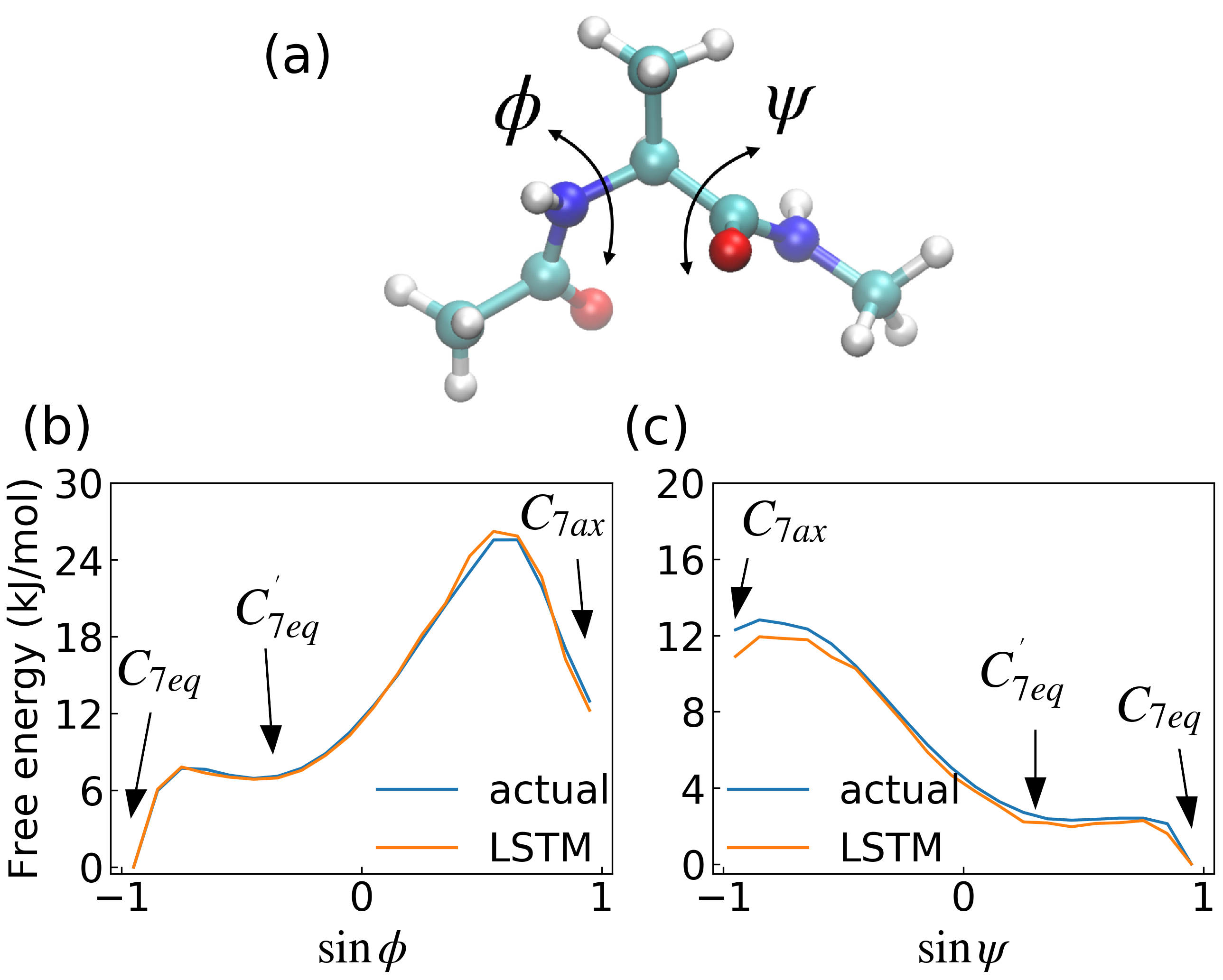}
  \caption
  {(a) The molecular structure of alanine dipeptide used in the actual MD simulatoin. The torsional angles $\phi$ and $\psi$ as the collective variables (CVs) are shown. (b) and (c) The 1-dimensional free energy curves along $\sin\phi$ and $\sin\psi$ are calculated using actual MD data and the data generated from LSTM.
}\label{fig:Boltzmann_aladip}
\end{figure}

Finally, we apply our LSTM model to the study of conformational transitions in alanine dipeptide, a model biomolecular system comprising 22 atoms, experiencing thermal fluctuations when coupled to a heat bath. The  structure of alanine dipeptide is shown in Fig. \ref{fig:Boltzmann_aladip}(a). While the full system comprises around 63 degrees of freedom, typically the torsional angles $\phi$ and $\psi$ are used to identify the conformations of this peptide. Over the years a large number of methods have been tested on this system in order to perform enhanced sampling of these torsions, as well as to construct optimal reaction coordinates.\cite{valsson2016enhancing,salvalaglio2014assessing,ma2005automatic,bolhuis2000reaction} Here we show that our LSTM model can very accurately capture the correct Boltzmann statistics as well as transition rates for moving between the two dominant metastable states known as $C_{7eq}$ and $C_{7ax}$. Importantly, the reconstruction of the equilibrium probability and transition kinetics, as shown in Fig. \ref{fig:Boltzmann_aladip} and Table \ref{tab:aladip_rate} is extremely accurate irrespective of the choice of one-dimensional projection time series fed into the LSTM. Specifically, we do this along $\sin\phi$ and $\sin\psi$, both of which are known to quite distant from an optimized kinetically truthful reaction coordinate\cite{smith2018multi,wang2019past}, where again we have excellent agreement between input and LSTM-predicted results.

\begin{table}
\renewcommand{\arraystretch}{1.5}
\begin{tabular}{|c|c|c|c|}
\hline
\multicolumn{4}{|c|}{Alanine dipeptide} \\
\hline
CVs & Label & $C_{7eq}$ to $C_{7ax}$ ($ps$) & $C_{7ax}$ to $C_{7eq}$ ($ps$) \\ \hline
\multirow{2}{*}{$\sin\phi$} 
& actual & 5689.22 $\pm$ 962.366 & 107.93 $ \pm$ 11.267 \\ \cline{2-4}
& LSTM & 5752.16 $\pm$ 710.399 & 103.81 $\pm$ 14.268 \\ \hline
\multirow{2}{*}{$\sin\psi$} 
& actual & 5001.42 $\pm$ 643.943 & 105.70$ \pm$ 13.521 \\  \cline{2-4}
& LSTM & 4325.01 $\pm$ 526.293 & 81.68 $\pm$ 10.288 \\ \hline
\end{tabular}
\caption{Inverse of transition rates for conformational transitions in alanine dipetide calculated from actual MD trajectories of LSTM model. Here we show the calculation along two different CVs: $\sin\phi$ and $\sin\psi$.
}
\label{tab:aladip_rate}
\end{table}

\subsection{Learning from single molecule force spectroscopy trajectory}
\label{sec:sm_fret}
\begin{figure}[!t]
  \centering
  \includegraphics[width=8cm]{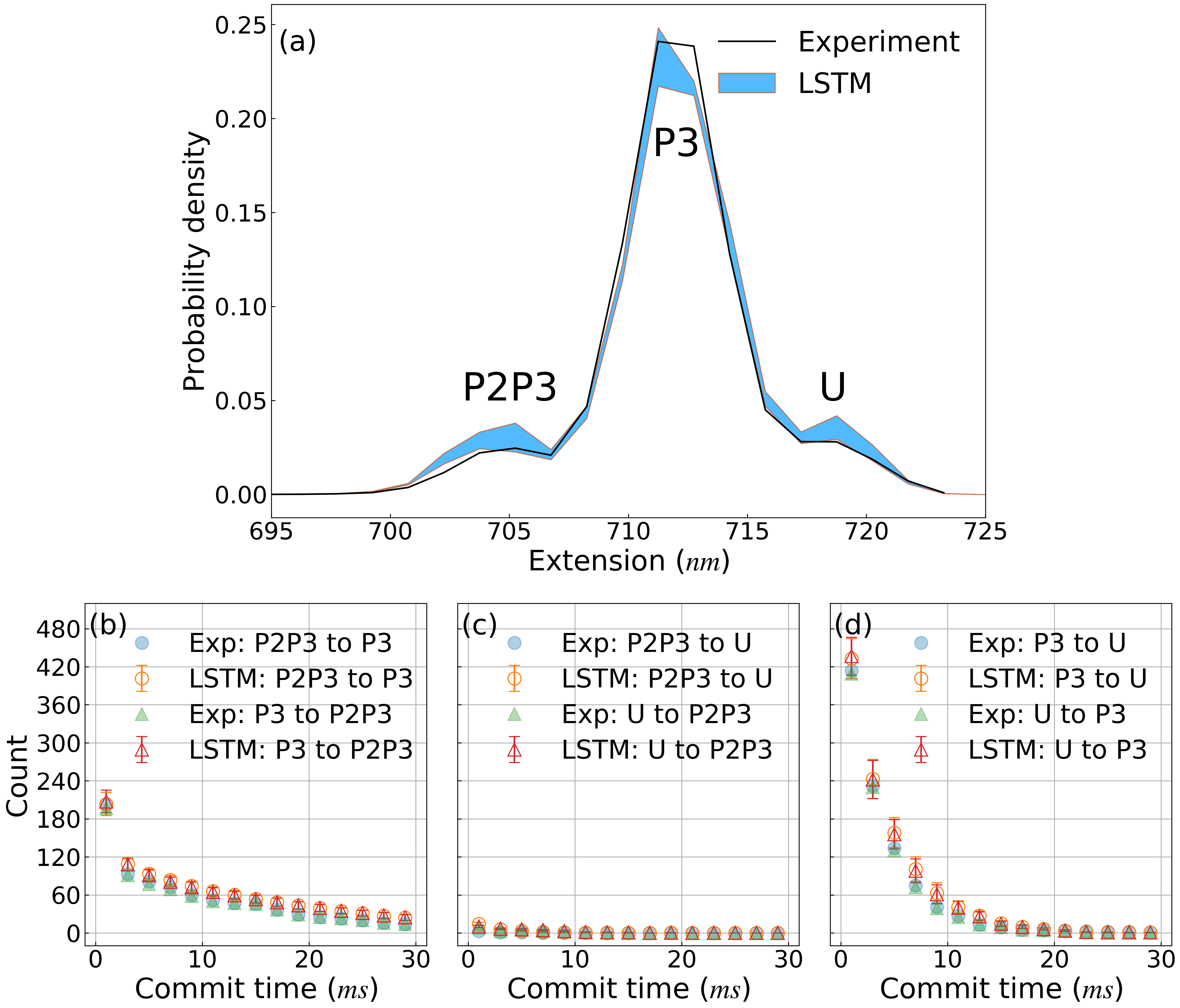}
  \caption
  {\one{Using LSTM model to learn thermodynamics and kinetics from a folding and unfolding trajectory taken from a single molecule force spectroscopy measurement:\cite{neupane2011single} (a) Comparison between the probability density learned by the LSTM model and calculated from the experimental data; (b)-(d) Commit time plots calculated by counting the transitions in the trajectory generated by LSTM and the experimental trajectory. The commit time as defined in Sec. \ref{sec:model_pots} is the minimum time that must be spent by the trajectory in a given state to be classified as having committed to it.}
}\label{fig:FRET_lstm}
\end{figure}

\one{In this section, we use our LSTM model to learn from single molecule force spectroscopy experiments of a multi-state riboswitch performed with a constant force of 10.9 $pN$. The data points are measured at 10 kHz (i.e., every 100 $\mu s$). Other details of the experiments can be found in Ref. \cite{neupane2011single}. The trajectory for a wide range of extensions starting 685 $nm$ up to 735 $nm$ was first spatially discretized into 34 labels, and then converted to a time series of one hot vectors, before being fed into the LSTM model. The results are shown in Fig. \ref{fig:FRET_lstm}. In Fig. \ref{fig:FRET_lstm} (a), we have shown an agreement between a profile of probability density averaged over 5
 independent training sets with the probability density calculated from the experimental data. Starting from the highest extension, the states are fully unfolded (U), longer intermediate (P3) and shorter intermediate (P2P3). \cite{neupane2011single}. From Fig. \ref{fig:FRET_lstm} (b)-(c), we see that the LSTM model captures the kinetics for moving between all 3 pairs of states for a very wide range of commitment times.}
 
\subsection{Embedding layer based kinetic distance}
\label{sec:embedding}
\begin{figure}[!h]
  \centering
  \includegraphics[width=8cm]{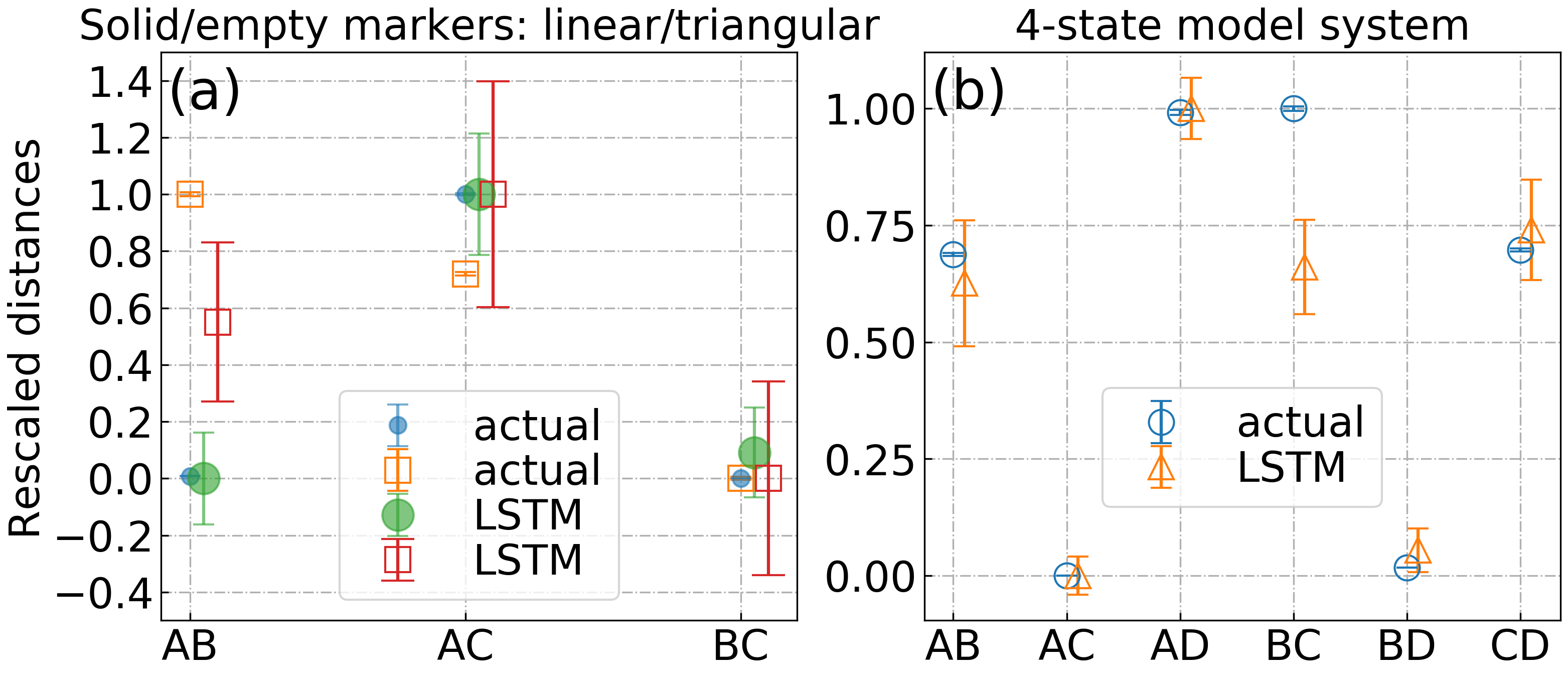}
  \caption
  {Our analysis of the embedding layer constructed for (a) the linear and triangular 3-state and (b) the 4-state model systems. In (a), we use solid circle and empty square markers respectively to represent linear and triangular 3-state model potentials. In each plot, the data points are shifted slightly to the right for clarity. The distances marked ``actual" and ``LSTM" represent rescaled mean transition times as per Eqs. \ref{eq:mean_trans_t} and \ref{eq:interconv_prob} respectively. Error bars were calculated over 50 different trajectories.
}\label{Embedding_3s_4s}
\end{figure}

In Sec. \ref{sec:embedding_kinetic}, we first derived a non-tractable relation for conditional transition probability in the embedding layer (Eq. \ref{eq:skipgramlikeQ}), and then through Eq. \ref{eq:conditional_p_fin} we introduced a tractable \textit{ansatz} in the spirit of Eq. \ref{eq:skipgramlikeQ}. In this section we revisit and numerically validate Eq. \ref{eq:conditional_p_fin}. Specifically, given any two embedding vectors $\mathbf{x}_l$ and $\mathbf{x}_m$ calculated from any two states $l$ and $m$, we estimate the conditional probability $Q_{lm}$ using Eq. \ref{eq:conditional_p_fin}. We use $Q_i$ to denotes the Boltzmann probability predicted by the LSTM model. We then write down the interconversion probability $k_{lm}$ between states $l$ and $m$ as:
\begin{align}
    k_{lm} = Q_lQ_{lm}+Q_mQ_{ml} \equiv 1/t_{lm}
    \label{eq:interconv_prob}
\end{align}
From inverting this rate we then calculate an LSTM-kinetic time as $t_{lm}\equiv {1 / k_{lm}} = 1/(Q_lQ_{lm}+Q_mQ_{ml})$.
In Fig. \ref{Embedding_3s_4s}, we compare $t_{lm}$ with the actual transition time  $\tau_{lm}$ obtained from the input data, defined as
\begin{align}
    \tau_{lm}&=T/\langle N_{lm}\rangle
    \label{eq:mean_trans_t}
\end{align}
Here $N_{lm}$ is the mean number of transitions between state $l$ and $m$. As this number varies with the precise value of commitment time, we average $N_{lm}$ over all commit times to get $\langle N_{lm}\rangle$. These two timescales $t_{lm}$ and $\tau_{lm}$ thus represent the average commute time or kinetic distance\cite{noe2016commute,noe2015kinetic} between two states $l$ and $m$. To facilitate the comparison between these two very differently derived timescales or kinetic distances, we rescale and shift them to lie between 0 and 1. The results in Fig. \ref{Embedding_3s_4s} show that the embedding vectors display the connectivity corresponding to the original high-dimensional configuration space rather than those corresponding to the one-dimensional projection. The model captures the correct connectivity by learning kinetics, which is clear evidence that it is able to bypass the projection error along any degree of freedom. The result also explains how is it that no matter what degree of freedom we use, our LSTM model still gives correct transition times. As long as the degree of freedom we choose to train the model can be used to discern all metastable states, we can even use Eq. \ref{eq:conditional_p_fin} to see the underlying connectivity. Therefore, the embedding vectors in LSTM can define a useful distance metric which can be used to understand and model dynamics, and are possibly part of the reason why LSTMs can model kinetics accurately inspite of quality of projection and associated non-Markvoian effects.

\subsection{Comparing LSTM with Markov state model and Hidden Markov Model}
\label{sec:Compare_MSM_HMM}

\begin{figure}[!h]
  \centering
  \includegraphics[width=8cm]{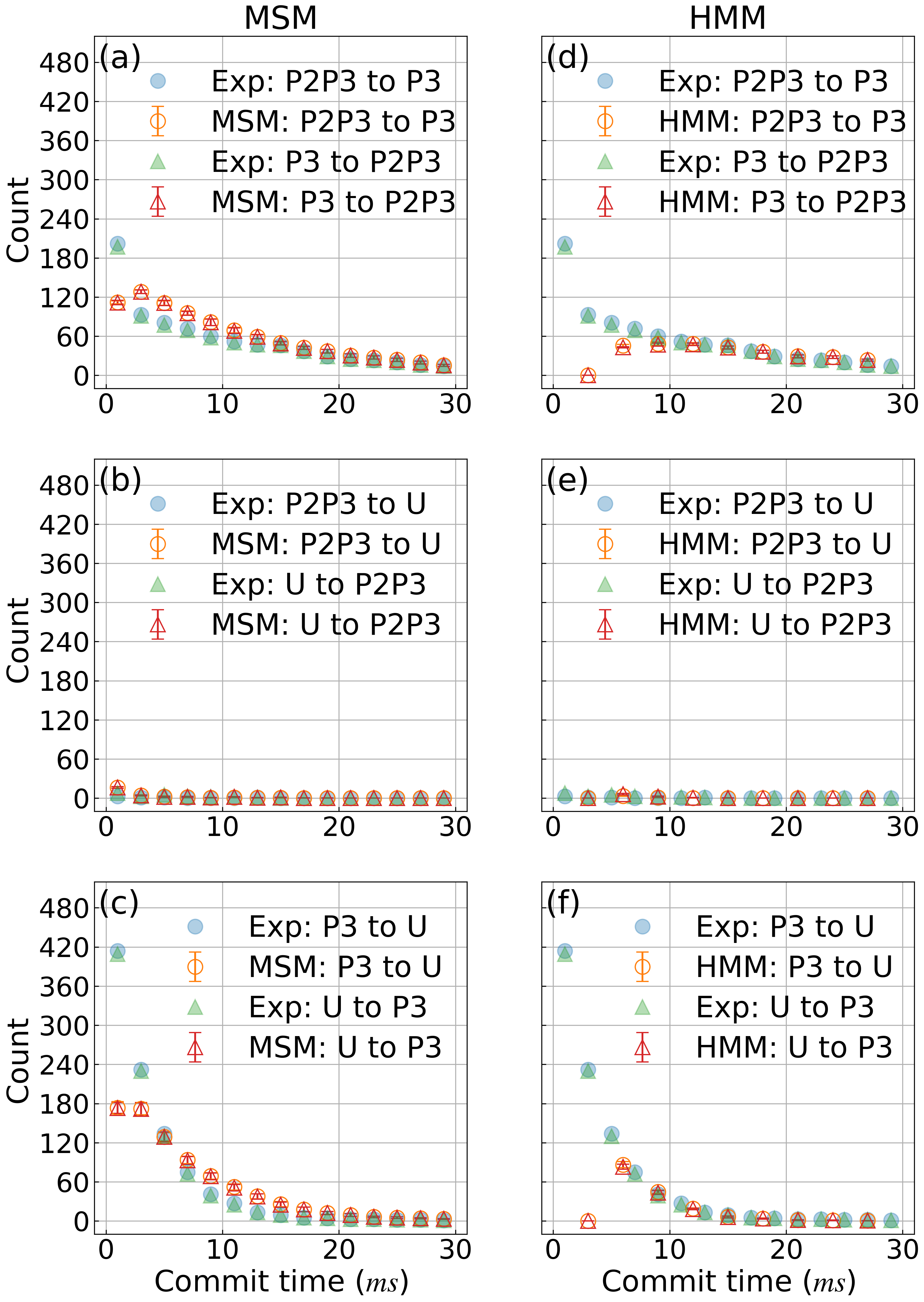}
  \caption
  { \one{Number of transitions between different pairs of metastable states as a function of commitment time defined in Sec. \ref{sec:model_pots} for the single molecule spectroscopy trajectory as learned by MSM (left column) and HMM (right column). Associated error bars are also provided.}
}\label{fig:FRET_k_msm_hmm_mix}
\end{figure}

\two{In this section, we briefly compare our LSTM model with standard approaches for building kinetic models from trajectories, namely the Markov state model (MSM)\cite{husic2018markov} and Hidden Markov model (HMM).\cite{eddy2004hidden,mckinney2006analysis,blanco2010analysis} Compared to LSTM, the MSM and HMM have smaller number of parameters, making them faster and more stable for simpler systems. However, both MSM and HMM require choosing an appropriate number of states and lag time\cite{bowman2009progress,husic2018markov,blanco2010analysis}. Large number of pre-selected states or small lag time can lead to non-Markovian behavior and result in an incorrect prediction. Even more critically, choosing a large lag time also sacrifices the temporal precision. On the other hand, there is no need to determine the lag time and number of states using the LSTM network because LSTM does not rely on the Markov property. Choosing hyperparameters such as $M$ and $L$ may be comparable to choosing number of hidden states for HMM, while very similar values of $M$ and $L$ worked for systems as different as MD trajectory of alanine dipeptide and single molecule force spectroscopy trajectory of a riboswitch. At the same time, LSTM always generates the data points with the same temporal precision as it has in the training data irrespective of the intrinsic timescales it learns from the system. In Fig. \ref{fig:FRET_k_msm_hmm_mix}, we provide the results of using HMM and MSM for the riboswitch trajectory, to be contrasted with similar plots using LSTM in Fig. \ref{fig:FRET_lstm}.  Indeed both MSM and HMM achieve decent agreement with the true kinetics only if the commit time is increased approximately beyond 10 ms, while LSTM as shown in Fig. \ref{fig:FRET_lstm} achieved perfect agreement for all commit times. From this figure, it can be seen that the LSTM model achieves an expected agreement with as fine of a  temporal precision as desired, even though we use 20 labels for alanine dipeptide and 34 labels for experimental data to represent the states. The computational efforts needed for the various approaches (LSTM, MSM and HMM) are also provided in the SI, where it can be seen that LSTM takes similar amount of effort as HMM. The package we used to build the MSM and HMM is PyEMMA with version 2.5.6.\cite{scherer2015pyemma}  The models were built with lag time=0.5$ms$ for MSM and lag time=3$ms$ for HMM, where the HMM were built with number of hidden states=3. A more careful comparison of the results along with analyses with other parameter choices such as different number of hidden states for HMM are provided in the SI, where we find all of these trends to persist.}

\section{Conclusions}
\label{se:conclusion}
In summary we believe this work demonstrates potential for using AI approaches developed for natural language processing such as speech recognition and machine translation, in unrelated domains such as chemical and biological physics. This work represents a first step in this direction, wherein we used AI, specifically LSTM flavor of recurrent neural networks, to perform kinetic reconstruction tasks that other methods \cite{perez2013identification,chodera2014markov} could have also performed. We would like to argue that demonstrating the ability of AI approaches to perform tasks that one could have done otherwise is a crucial first step. In future works we will exploring different directions in which the AI protocol developed here could be used to perform tasks which were increasingly non-trivial in non-AI setups.
More specifically, in this work we have shown that a simple character-level language model based on LSTM neural network can learn a probabilistic model of a time series generated from a physical system such as an evolution of Langevin dynamics or MD simulation of complex molecular models. We show that the probabilistic model can not only learn the Boltzmann statistics but also capture a large spectrum of kinetics. The embedding layer which is designed for encoding the contextual meaning of words and characters displays a nontrivial connectivity and has been shown to correlate with the kinetic map defined for reversible Markov chains.\cite{mikolov2013distributed,noe2015kinetic,noe2016commute} For different model systems considered here, we could obtain correct timescales and rate constants irrespective of the quality of order parameter fed into the LSTM. As a result, we believe this kind of model outperforms traditional approaches for learning thermodynamics and kinetics, which can often be very sensitive to the choice of projection. Finally, the embedding layer can be used to define a new type of distance metric for high-dimensional data when one has access to only some low-dimensional projection. We hope that this work represents a first step in the use of RNNs for modeling, understanding and predicting the dynamics of complex stochastic systems found in biology, chemistry and physics.


\section{Data Availability}
\label{sec:data_avail}
The single-molecule force spectroscopy experiment data for riboswitch was obtained from the authors of Ref. \cite{neupane2011single} and they can be contacted for the same. All the other data associated with this work is available from the corresponding author on request.

\section{Code Availability}
\label{sec:code_avail}
MSM and HMM analyses were conducted with PyEMMA version 2.5.6.\cite{scherer2015pyemma} and available at http://www.pyemma.org. A Python based code of the LSTM language model introduced in Sec. \ref{sec:computational_results} is implemented using keras\cite{Charles2013} with tensorflow-gpu\cite{abadi2019tensorflow} as a backend, and available for public use at https://github.com/tiwarylab/LSTM-predict-MD.

\section*{Acknowledgments}
P.T. thanks Dr. Steve Demers for suggesting the use of
LSTMs. The authors thank Carlos Cuellar for the help in early stages of this project, \one{Michael Woodside for sharing the single molecule trajectory with us,} Yihang Wang for in-depth discussions, Dedi Wang, Yixu Wang, Zachary Smith for their helpful insights and suggestions. Acknowledgment is made to the Donors of the American Chemical Society Petroleum Research Fund for partial support of this research (PRF 60512-DNI6). We also thank Deepthought2, MARCC and XSEDE (projects CHE180007P and CHE180027P) for computational resources used in this work.

\end{document}